\documentclass[aps,prd,floats,floatfix,nofootinbib,preprintnumbers,superscriptaddress]{revtex4-1}
\usepackage{graphicx,amsmath}
\usepackage{amssymb}
\usepackage{amsfonts}
\usepackage{hyperref}
\usepackage[bottom]{footmisc}

\begin{document}

\preprint{NUHEP-TH/12-011}

\title{Can the 126~GeV boson be a pseudoscalar?}

\author{Baradhwaj Coleppa}
\email {barath@physics.carleton.ca}

\affiliation {Ottawa-Carleton Institute for Physics,
Carleton University,
Ottawa, Ontario K1S 5B6, Canada}

\author{Kunal Kumar}
\email {kkumar@u.northwestern.edu }

\affiliation{Northwestern University, 2145 Sheridan Road, Evanston, Illinois 60208, USA}

\author{Heather E.\ Logan}
\email {logan@physics.carleton.ca }

\affiliation {Ottawa-Carleton Institute for Physics,
Carleton University,
Ottawa, Ontario K1S 5B6, Canada}

\date{August 13, 2012}

\begin{abstract}
We test the possibility that the newly-discovered 126~GeV boson is a pseudoscalar by examining the correlations among the loop-induced pseudoscalar decay branching fractions to $\gamma\gamma$, $ZZ^*$, $Z\gamma$, and $WW^*$ final states in a model-independent way.  These four decays are controlled by only two effective operators, so that the rates in $Z\gamma$ and $WW^*$ are predicted now that the rates in $\gamma\gamma$ and $ZZ^*,Z\gamma^* \to 4 \ell$ have been measured.  We find that the pseudoscalar possibility is disfavored but not conclusively excluded.  Experimental exclusion of the $Z\gamma$ decay to well below $\sigma/\sigma_{\rm SM} \sim 170$ or conclusive observation of the $WW^*$ decay near the Standard Model rate would eliminate the pseudoscalar possibility.  The $Z\gamma$ exclusion should be possible using existing data.  The only loophole in our argument is the possibility that the $4\ell$ signal comes from pseudoscalar decays to a pair of new neutral gauge bosons with mass near the $Z$ pole.
\end{abstract}

\maketitle
%%%%%%%%%%%%%%%%%%%%%%%%%%%%%%%%%%%%%%%%%%%%%%
\section{Introduction}

The ATLAS and CMS experiments at the CERN Large Hadron Collider (LHC) have discovered a new boson with mass of about 126~GeV~\cite{ATLASdiscovery,CMSdiscovery} using about 5~fb$^{-1}$ of integrated luminosity per experiment at each of 7~TeV and 8~TeV.  Decays into $\gamma\gamma$ and $ZZ^* \to 4\ell$ final states have been observed with significances of 4.1--4.5$\sigma$~\cite{CMSdiscovery,ATLASdiphoton} and 3.2--3.4$\sigma$~\cite{CMSdiscovery,ATLASZZ}, respectively, at each experiment.  These final states have good mass resolution of order 1~GeV.  Decays into $WW^* \to \ell\nu\ell\nu$ have also been observed at the 1.6--2.8$\sigma$ level~\cite{CMSdiscovery,ATLASWW}, albeit with much poorer mass resolution.  Searches for decays into $\tau\tau$ and $b \bar b$ final states have so far been inconclusive~\cite{CMSdiscovery,ATLAStautau,ATLASbb}.  The observed signal rates in all channels are consistent with the new boson being the Standard Model (SM) Higgs boson.

A key question posed by this discovery is the determination of the spin and CP quantum numbers of the new boson.  Observation of the two-photon final state excludes the possibility of spin 1 via the Landau-Yang theorem~\cite{LandauYang}.\footnote{A spin-1 boson could still be accommodated in the scenario that the $\gamma\gamma$ signal comes from decays to two very light intermediary particles which each decay to a pair of tightly collimated photons~\cite{Draper:2012xt}.}  Spin-2 can be distinguished from spin-0 by examining the angular distributions of the final-state photons~\cite{Gao:2010qx,Ellis:2012wg}, the leptons in $ZZ^* \to 4 \ell$~\cite{Choi:2002jk,Djouadi:2005gi,Gao:2010qx,DeRujula:2010ys,DeSanctis:2011yc} and $WW^* \to \ell \nu \ell \nu$~\cite{Ellis:2012wg} final states, and the angular correlations between the jets produced in vector boson fusion~\cite{Hagiwara:2009wt}.

A spin-0 resonance can have CP $= +1$ (scalar, like the SM Higgs) or CP $= -1$ (pseudoscalar).  The CP nature of a spin-0 resonance can be determined using the angular distributions of the leptons in $ZZ^* \to 4 \ell$~\cite{Dell'Aquila:1985ve,Barger:1993wt,Chang:1993jy,Soni:1993jc,Arens:1994wd,Choi:2002jk,Djouadi:2005gi,Cao:2009ah}, the angular distributions of the jets produced in vector boson fusion~\cite{Plehn:2001nj} or $gg \to Xjj$~\cite{Campanario:2010mi}, hadronic event shape observables~\cite{Englert:2012ct}, or the spin correlations in decays to $\tau\tau$~\cite{Berge:2008dr}.
All of these measurements require more integrated luminosity than has been used up to now for Higgs analyses.  

Several pseudoscalar interpretations of the 126~GeV excess, particularly in the $\gamma\gamma$ final state, have been put forward in the context of specific models~\cite{Bernreuther:2010uw,Burdman:2011ki,Holdom:2012pw,Cervero:2012cx,Frandsen:2012rj,Moffat:2012ix,Chivukula:2012cp}.  A scalar-pseudoscalar admixture has also been considered in Ref.~\cite{Barroso:2012wz}.

In this paper we attempt to constrain the pseudoscalar possibility in a model-independent way by examining the correlations among the $\gamma\gamma$, $ZZ^*$, $WW^*$, and $Z\gamma$ decay modes.  For a pseudoscalar, all of these decays arise from loop-induced effective couplings, which can be parameterized in the electroweak basis in terms of just two operator coefficients.
A similar approach has been taken to constrain an electroweak-singlet CP-even scalar that has only loop-induced couplings to $WW$ and $ZZ$ in Ref.~\cite{Low:2012rj}.  We improve upon the analysis in Ref.~\cite{Low:2012rj} by taking into account the very large contribution to the $4\ell$ final state coming from $\phi \to Z \gamma^* \to 4 \ell$.  

After fitting the ratio of the $\gamma\gamma$ and $4\ell$ final states to the SM Higgs prediction, we find that the branching fraction for the pseudoscalar into $WW^*$ is dramatically suppressed compared to the SM Higgs and the branching fraction into $Z\gamma$ is dramatically enhanced. The suppression of the $WW^*$ mode disfavors the pseudoscalar at somewhat less than the 3$\sigma$ level, based on the ATLAS excess in this channel~\cite{ATLASWW}.  The enhancement of the $Z\gamma$ mode is not large enough to allow the pseudoscalar to be excluded based on the LHC measurements of the continuum $Z\gamma$ production cross section; however, it is large enough that a resonance search should easily have the sensitivity to exclude the pseudoscalar possibility.

We also consider possible loopholes in the argument for the large and easily-excludable enhancement of the $Z\gamma$ branching fraction.  We conclude that the only way out is if the observed $4\ell$ final state comes from decays of the pseudoscalar into a pair of new neutral gauge bosons with mass near the $Z$ pole, which then decay to $\ell^+\ell^-$.  The strong suppression of the $WW^*$ branching fraction survives as a distinguishing feature in this case.

This paper is organized as follows.  In Sec.~\ref{sec:couplings} we define the effective operator parameterization for the pseudoscalar couplings to gauge bosons.  In Sec.~\ref{sec:b} we determine the parameter value needed to match the observed rates of the 126~GeV boson into $\gamma\gamma$ and $4\ell$ final states, taking care to properly include contributions from $\phi \to Z \gamma^* \to 4\ell$.  In Sec.~\ref{sec:predictions} we compute the resulting predictions for $\phi \to WW^*$ and $\phi \to Z\gamma$ and discuss how they can be used to confirm or exclude the pseudoscalar possibility.  In Sec.~\ref{sec:loopholes} we consider potential loopholes in our analysis.  We conclude in Sec.~\ref{sec:conclusions}.

%%%%%%%%%%%%%%%%%%%%%%%%%%%%%%%%%%%%%%%%%%%
\section{Pseudoscalar couplings}
\label{sec:couplings}

A pseudoscalar has no gauge-invariant renormalizable couplings to SM gauge fields.  Instead, we model the interaction of the pseudoscalar with the SM gauge fields using the following effective Lagrangian:
%%%%%%%%%%%%
\begin{equation}
 \mathcal{L}=c\frac{\alpha_s}{4\pi v}\phi G^{a}_{\mu\nu}\widetilde{G}^{a\mu\nu}+a\frac{\alpha}{4\pi v}\phi\left[B_{\mu\nu}\widetilde{B}^{\mu\nu}+bW^{i}_{\mu\nu}\widetilde{W}^{i\mu\nu} \right].
\label{eqn:phi-L}
\end{equation}
Here $v = 246$~GeV is the SM Higgs vacuum expectation value.  The dual field strength tensors are defined by $\widetilde{G}^{a\mu\nu}=\epsilon^{\mu\nu\rho\sigma}G^a_{\rho\sigma}$, and similarly for the other field strength tensors.  The above Lagrangian allows us to parameterize the decays of $\phi$ to $\gamma\gamma$, $W^{+}W^{-}$, $ZZ$, $Z\gamma$, and $gg$. The amplitudes for these processes can be easily derived from Eq.~(\ref{eqn:phi-L}).  At leading order, their dependence on the parameter $b$ is:
\begin{eqnarray}
 \mathcal{M}\left(\phi\rightarrow WW \right)&\propto&b \nonumber \\
 \mathcal{M}\left(\phi\rightarrow ZZ \right)&\propto&(b\,\textrm{cos}^2\theta_{W}+\textrm{sin}^2\theta_{W}) \nonumber \\
 \mathcal{M}\left(\phi\rightarrow \gamma\gamma \right)&\propto&(\textrm{cos}^2\theta_{W}+b\,\textrm{sin}^2\theta_{W}) \nonumber \\
 \mathcal{M}\left(\phi\rightarrow Z\gamma \right)&\propto&(b-1),
\label{eqn:amp}
\end{eqnarray}
where $\theta_W$ is the weak mixing angle.
%\begin{eqnarray}
% \mathcal{M}\left(\phi\rightarrow WW \right)&=&\frac{\sqrt{2}b\alpha}{\pi v}M_{\phi}\left(M_{\phi}^2-4M_{W}^2 \right)^{1/2} \nonumber \\
% \mathcal{M}\left(\phi\rightarrow ZZ \right)&=&\frac{(b\,\textrm{cos}^2\theta_{w}+\textrm{sin}^2\theta_{w})\alpha}{\pi v}M_{\phi}\left(M_{\phi}^2-4M_{Z}^2 \right)^{1/2} \nonumber \\
% \mathcal{M}\left(\phi\rightarrow \gamma\gamma \right)&=&\frac{(\textrm{cos}^2\theta_{w}+b\,\textrm{sin}^2\theta_{w})\alpha}{\pi v}M_{\phi}^2 \nonumber \\
 %\mathcal{M}\left(\phi\rightarrow Z\gamma \right)&=&\frac{(b-1)\textrm{cos}\,\theta_{w}\textrm{sin}\,\theta_{w}\alpha}{\pi v}\left(M_{\phi}^4-2M_{\phi}^2M_{Z}^2+M_{Z}^4 \right)^{1/2} 
%\label{eqn:amp}
%\end{eqnarray}

\begin{figure}
\resizebox{0.6\textwidth}{!}{\includegraphics{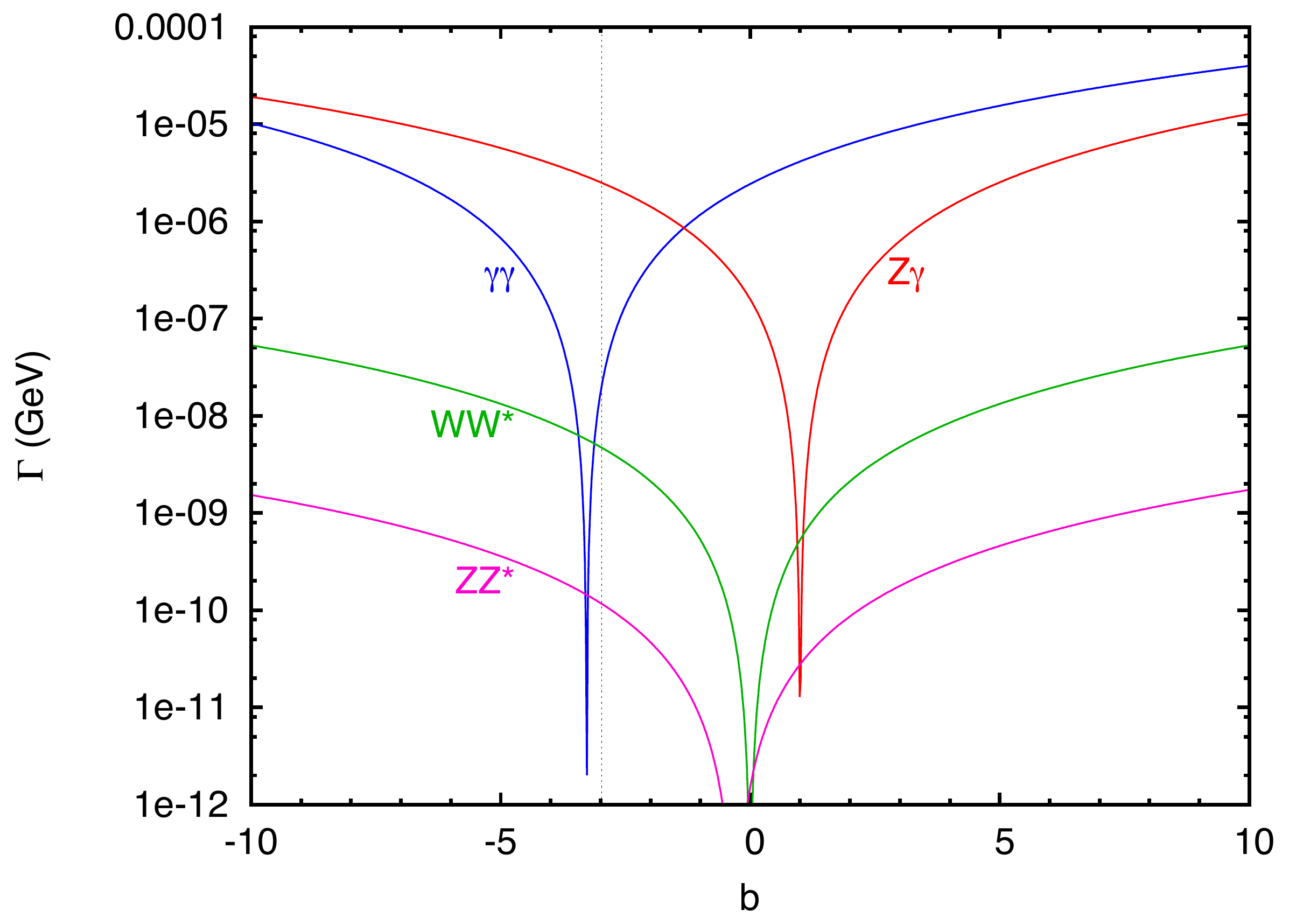}}
\caption{Theoretical prediction for the partial widths of $\phi$ to $\gamma\gamma$ (blue), $Z\gamma$ (red), $WW^{*}$ (green), and $ZZ^{*}$ (magenta) as a function of $b$, for $M_{\phi} = 126$~GeV and $a = 1$.  The normalization on the $y$-axis can be changed by varying $a$; all the partial widths shown scale with $a^2$.  The vertical dotted line shows the value of $b$ that we will use to match the SM rates in $\gamma\gamma$ and $4\ell$.}
\label{fig:phi-BR}
\end{figure}

It is readily apparent from the above equation that the decay amplitude to a particular gauge boson pair can be adjusted by tuning the parameter $b$. In particular, each amplitude vanishes at a different value of $b$:
\begin{eqnarray}
	\mathcal{M}\left(\phi\rightarrow WW \right) = 0&& \ \ {\rm when} \ b = 0 \nonumber \\
 	\mathcal{M}\left(\phi\rightarrow ZZ \right) = 0&& \ \ {\rm when} \ b = - \tan^2\theta_W \nonumber \\
 	\mathcal{M}\left(\phi\rightarrow \gamma\gamma \right) = 0&& \ \ {\rm when} \ b = - \cot^2\theta_W \nonumber \\
	 \mathcal{M}\left(\phi\rightarrow Z\gamma \right) = 0&& \ \ {\rm when} \ b = 1.
\label{eqn:ampzeros}
\end{eqnarray}
We illustrate this behavior by plotting the partial widths of $\phi$ to $Z\gamma$, $\gamma\gamma$, $ZZ^*$, and $WW^*$ as a function of $b$ in Fig.~\ref{fig:phi-BR}, for a $\phi$ mass of 126~GeV.

Notice in particular that the partial widths to $WW^*$ and $ZZ^*$ are in general much smaller than those to $\gamma\gamma$ or $Z\gamma$, except for $b$ values near the zeros of the $\gamma\gamma$ or $Z\gamma$ amplitudes.  This is due to the kinematic suppression of these decays for $\phi$ masses below the $WW$ and $ZZ$ thresholds.  This kinematic suppression will force us to choose $b$ very close to the zero of the $\gamma\gamma$ amplitude in order to match the rates in $\gamma\gamma$ and $4\ell$ experimentally observed for the 126~GeV boson.

%%%%%%%%%%%%%%%%%%%%%%%%%%%%%%%%%%%%%%%
\section{Matching the observed rates}
\label{sec:b}

We now determine the value of $b$ required to fit the observed rates in $\gamma\gamma$ and $4\ell$ final states of the 126~GeV boson.  This can be done using only the ratio of rates (the overall production rate for $gg \to \phi$ can be adjusted by varying $c$).  The fit is complicated by the fact that $\phi \to Z \gamma^* \to 4 \ell$ contributes substantially to the $4\ell$ final state selected by the LHC experiments.

We proceed as follows.  In the SM, the ratio of Higgs partial widths into $ZZ^*$ versus $\gamma\gamma$ for a Higgs mass of 126~GeV is predicted to be~\cite{Dittmaier:2012vm}\footnote{The contribution of SM $H \to Z\gamma^*$ to the $4\ell$ final state is experimentally negligible because $\Gamma(H \to Z\gamma)/\Gamma(H \to ZZ^*) = 0.056$; taking the photon off shell suppresses the $Z\gamma^*$ contribution even further.}
\begin{equation}
	R_{ZZ/\gamma\gamma}^{\rm SM} \equiv \frac{\Gamma(H \to ZZ^*)}{\Gamma(H \to \gamma\gamma)} \simeq 12.7.
\end{equation}
This can be rewritten in terms of a partial width after cuts in the $4e$ final state as
\begin{equation}
	R_{ZZ/\gamma\gamma}^{\rm SM} = \frac{\Gamma(H \to ZZ^* \to 4e)}{\Gamma^c(H \to ZZ^* \to 4e)} \times
	\frac{1}{\left[ {\rm BR}(Z \to ee) \right]^2} \times
	\frac{\Gamma^c(H \to ZZ^* \to 4e)}{\Gamma(H \to \gamma\gamma)},
	\label{eq:Rfactored}
\end{equation}
where $\Gamma^c$ refers to the width after cuts (defined in terms of the total signal rate after cuts; the boson production cross section and total width factors cancel in the ratio with $\Gamma(H \to \gamma\gamma)$).  The first term in Eq.~(\ref{eq:Rfactored}) is determined by a Monte Carlo simulation using CalcHEP~\cite{CalcHEP}, the second is computed at leading order to match the signal calculation, and the third is what has been measured by ATLAS and CMS.

We match the observed rates by requiring that the ratio of rates to $4\ell$ and $\gamma\gamma$ for the pseudoscalar be the same as that of the SM Higgs; i.e., in Eq.~(\ref{eq:Rfactored}) we make the replacement,
\begin{equation}
	\frac{\Gamma^c(H \to ZZ^* \to 4e)}{\Gamma(H \to \gamma\gamma)}
	\ \rightarrow \
	\frac{\Gamma^c(\phi \to 4e)}{\Gamma(\phi \to \gamma\gamma)}.
	\label{eq:replacement}
\end{equation}

\begin{figure}
\resizebox{\textwidth}{!}{\includegraphics{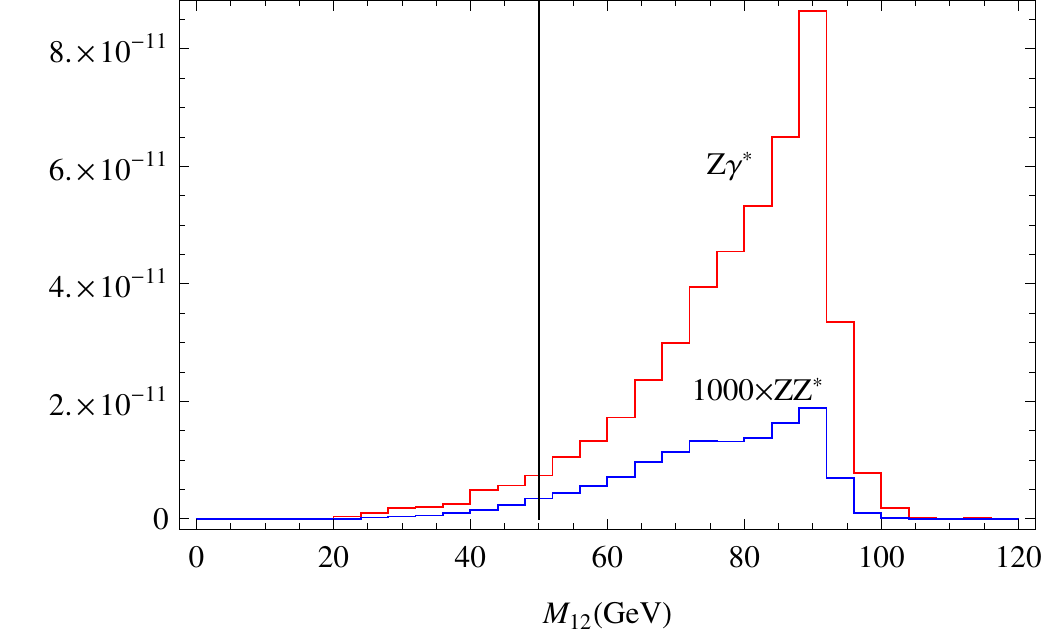}\includegraphics{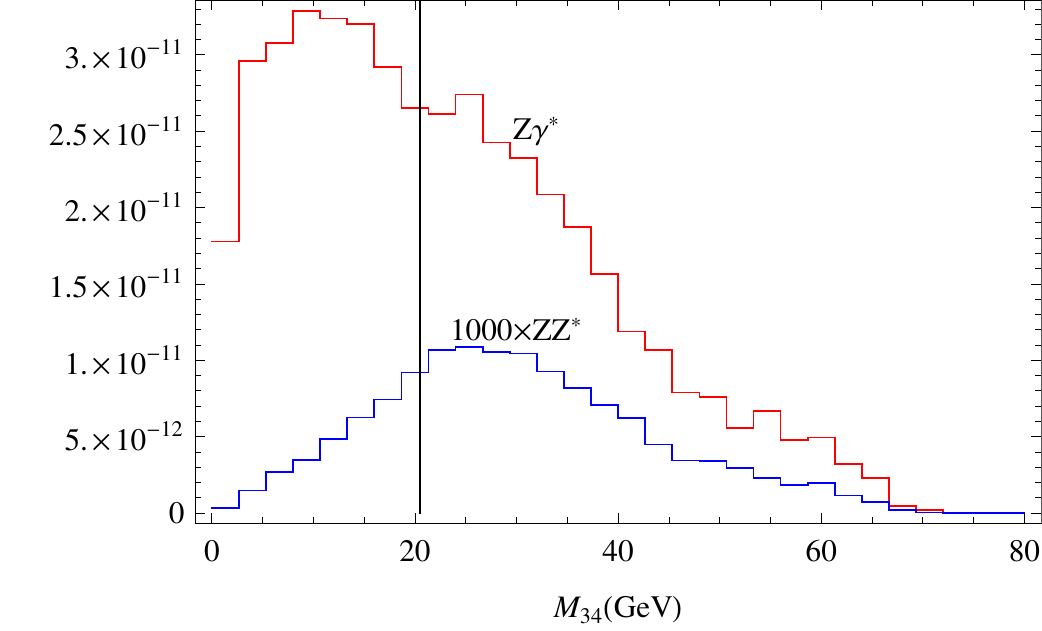}}
\caption{Invariant mass distributions $M_{12}$ (left) and $M_{34}$ (right) for the $4e$ final state event selection (see text for definitions), after application of the $\Delta R$ cut.  We show separately the contributions from $\phi \to Z\gamma^* \to 4e$ (red) and $\phi \to ZZ^* \to 4e$ (blue, multiplied by 1000).  The vertical lines indicate the lower cuts on $M_{12}$ at 50~GeV and on $M_{34}$ at 20.5~GeV.}
\label{fig:M34}
\end{figure}

To determine the acceptance for $H \to ZZ^* \to 4e$ and the partial width after cuts for $\phi \to 4e$, we apply a subset of the ATLAS selection cuts~\cite{ATLASZZ}:
\begin{itemize}
\item Each pair of leptons must be separated by at least $\Delta R = 0.1$.
\item The electron-positron pair with invariant mass $M_{12}$ closest to the $Z$ mass must satisfy 50~GeV $< M_{12} < 106$~GeV.
\item The remaining electron-positron pair must have an invariant mass $M_{34}$ between 20.5~GeV and 115~GeV.  This cut affects how much the process $Z \gamma^* \to 4e$ contributes to the signal after cuts, as can be seen in Fig.~\ref{fig:M34}.  Note that the $Z\gamma^*$ contribution after cuts is about 2400 times the size of the true $ZZ^*$ contribution, in extreme contrast to the SM Higgs. 
\end{itemize}

Putting everything together and adjusting the value of $b$ until Eq.~(\ref{eq:Rfactored}) is satisfied, we obtain two solutions for $b$, corresponding to the two sides of the zero in the $\gamma\gamma$ rate shown in Fig.~\ref{fig:phi-BR}.  For concreteness we take the side with smaller $|b|$,
\begin{equation}
	b = -\cot^2\theta_W (1 + \epsilon), \qquad {\rm where} \ \epsilon = -0.092.
\end{equation}
% b = -2.975, \cot^2\theta_W = 3.2775.

Note that $b$ must be tuned to within ten percent of $-\cot^2\theta_W$.  Because of the rapid variation of $\Gamma(\phi \to \gamma\gamma)$ in this range of $b$, factor-of-two changes in the ratio $\Gamma^c(\phi \to 4e)/\Gamma(\phi \to \gamma\gamma)$ result in only small absolute changes in the value of $b$.  Our results in the next section for the rates for $\phi \to WW^*$ and $\phi \to Z\gamma$ do not depend sensitively on such small variations in $b$.
This justifies the several assumptions that we have made in the above analysis:
\begin{itemize}

\item We fit the value of $b$ to the SM prediction for $\Gamma(H \to ZZ^*)/\Gamma(H \to \gamma\gamma)$, rather than to the measured ratio.  The measured ratio is smaller than the SM prediction, but consistent within the experimental uncertainties (ATLAS and CMS both measure a rate in the $\gamma\gamma$ channel somewhat larger than the SM expectation).  Fitting to the SM prediction rather than the measurements yields a slightly different value of $b$ but will not change our overall conclusions.

\item We assume that the experimental acceptance for the pseudoscalar $\phi \to \gamma\gamma$ signal is the same as that for the SM $H \to \gamma\gamma$ signal.  This allows us to write $\Gamma(H \to \gamma\gamma)$ in the denominator of Eq.~(\ref{eq:Rfactored}) and $\Gamma(\phi \to \gamma\gamma)$ in the denominator of Eq.~(\ref{eq:replacement}) without including selection efficiencies.  Because the decay kinematics are identical, the difference in efficiencies in this case can come only from the difference in the $p_T$ distribution of the initially produced boson and the kinematics of the accompanying jets.  A proper evaluation of the difference in efficiencies would require a full simulation of Higgs and pseudoscalar production, which is beyond the scope of our analysis.  In any case, this effect is small enough that it will not change our conclusions.

\item We do not apply $p_T$ or rapidity cuts to the leptons in the $4e$ final states.  This will affect our results only to the extent that $H$ and $\phi$ are produced with different transverse momentum distributions.  Again, this effect is small enough that it will not change our conclusions.

\end{itemize}

%%%%%%%%%%%%%%%%%%%%%%%%%%%%%%%%%%%%%%%%%%%%%%%
\section{Predictions for $WW^*$ and $Z\gamma$ final states}
\label{sec:predictions}

With $b$ in hand, we can read off the predictions for the pseudoscalar decay widths to $WW^*$ and $Z\gamma$.  We find,
\begin{eqnarray}
	R_{WW/\gamma\gamma}^{\phi} 
	&\equiv& \frac{\Gamma(\phi \to WW^*)}{\Gamma(\phi \to \gamma\gamma)} 
	= 0.229
	\nonumber \\
	R_{Z\gamma/\gamma\gamma}^{\phi} 
	&\equiv& \frac{\Gamma(\phi \to Z\gamma)}{\Gamma(\phi \to \gamma\gamma)}
	=  121.
\end{eqnarray}
For comparison, the corresponding ratios for the 126~GeV SM Higgs are~\cite{Dittmaier:2012vm}
\begin{equation}
	R_{WW/\gamma\gamma}^{\rm SM} \simeq 101, \qquad
	R_{Z\gamma/\gamma\gamma}^{\rm SM} \simeq 0.711.
\end{equation}

The smallness of the rate for $\phi \to WW^*$, 440 times smaller than the SM $H \to WW^*$ rate, disfavors the pseudoscalar insofar as ATLAS and CMS have observed weak signals in this channel ($1.6\sigma$ from CMS and $2.8\sigma$ from ATLAS~\cite{CMSdiscovery,ATLASWW}).  In particular, the ATLAS analysis of 8~TeV data, which provides the largest contribution to their overall significance, so far includes only the $e \mu$ final state.  This is interesting because $Z\gamma^* \to ee\nu\nu$, $\mu\mu\nu\nu$ will spill in to the same-flavor event selections, thereby boosting the rates in these channels, but it cannot contaminate the $e\nu\mu\nu$ sample.  

The most direct probe, however, is the very large enhancement of the rate for $pp \to \phi \to Z\gamma$, 170 times larger than the SM $pp \to H \to Z\gamma$ rate.  This translates~\cite{Dittmaier:2012vm} into a cross section for $pp \to \phi \to Z\gamma$ of 4.8~pb (6.0~pb) at 7~TeV (8~TeV).  This cross section is too small to be constrained by direct comparison with the continuum $Z\gamma$ cross section, measured by ATLAS and CMS to be about 290~pb (up to 10\% uncertainty) in the fiducial region of $E_T^{\gamma} > 10$~GeV, $\Delta R_{\ell, \gamma} > 0.7$, $M_{\ell\ell} > 50$~GeV~\cite{Aad:2011tc,Chatrchyan:2011rr,Aad:2012mr}.

On the other hand, $\phi \to Z\gamma \to \ell \ell \gamma$ produces an invariant mass peak.
A phenomenological study of this channel has been done for the SM Higgs in Ref.~\cite{Gainer:2011aa}, which found a 95\% CL exclusion reach of $\sigma/\sigma_{\rm SM} \simeq 4$ for a 126~GeV Higgs with 20~fb$^{-1}$ at 8~TeV after an optimized cut on a multivariate discriminant (including only statistical uncertainties).  Scaling to the luminosity of $\sim 5$~fb$^{-1}$ already used for Higgs analyses at 8~TeV, this should translate into an idealized exclusion reach of about $\sigma/\sigma_{\rm SM} \simeq 8$.  The pseudoscalar prediction of $\sigma/\sigma_{\rm SM} \simeq 170$ in the $Z\gamma$ channel should therefore be easy to either discover or exclude in current data, even with a non-optimized analysis.

Finally we comment on the total width of the pseudoscalar.  So long as BR($\phi \to gg) \ll 1$, the total width is dominated by $Z\gamma$ and we find BR($\phi \to \gamma\gamma) \simeq 1/121 = 0.83\% \simeq 3.6 \, {\rm BR}(H \to \gamma\gamma)$.  In this case the signal rate in $pp \to \phi \to \gamma\gamma$ matches the SM Higgs prediction for $\sigma(gg \to \phi) \simeq 0.3\, \sigma(gg \to H)$.  This can be used to fix $c$.  The overall scaling factor $a$ on the effective electroweak gauge couplings can then be adjusted to maintain BR($\phi \to gg) \ll 1$.\footnote{The parameters $c$ and $a$ can of course be increased together to maintain a constant value of $\sigma(pp \to \phi) \times {\rm BR}(\phi \to \gamma\gamma)$ while increasing the total width of $\phi$.}  Insofar as BR($H \to gg) = 0.085 \ll 1$ in the SM~\cite{Dittmaier:2012vm}, the total width of the pseudoscalar can therefore be comparable to the SM Higgs total width of about 4.2~MeV~\cite{Dittmaier:2012vm}, i.e., well below the detector resolution.

%%%%%%%%%%%%%%%%%%%%%%%%%%%%%%%%%%%%%%%
\section{Potential loopholes}
\label{sec:loopholes}

We can think of only two potential loopholes in our prediction for the easily-excludable large enhancement of the $Z\gamma$ rate for a pseudoscalar.  Both involve the breakdown of the effective Lagrangian description of Eq.~(\ref{eqn:phi-L}).

\subsection{Dependence of the effective couplings on gauge boson masses}

Starting from a renormalizable theory, the effective couplings of the pseudoscalar to a pair of gauge bosons in Eq.~(\ref{eqn:phi-L}) are generated by loops of fermions  that couple to the pseudoscalar and carry the appropriate gauge charge.  In the limit that the fermions are much heavier than the pseudoscalar, $W$, and $Z$, the effective Lagrangian is recovered.  However, if the fermions are light enough, the loop amplitude will depend on the invariant masses of the external gauge bosons.  In particular, the coefficients $a$ of the hypercharge effective coupling and $ab$ of the SU(2)$_L$ effective coupling could take different values for the $Z\gamma$ final state than for the $\gamma\gamma$ final state.  This has the effect of enhancing the contribution of any particular fermion to $Z\gamma$ compared to its contribution to $\gamma\gamma$, as follows.

The fermion loop function for on-shell $\phi \to \gamma\gamma$ is given by~\cite{HHG}
\begin{equation}
	F_{1/2}^{\phi}(\tau) = -2 \tau f(\tau) \rightarrow -2 \ {\rm for} \ m_f \to \infty,
\end{equation}
and for on-shell $\phi \to Z \gamma$ by
\begin{equation}
	I_2^{\phi}(\tau,\lambda) = - \frac{\tau\lambda}{2 (\tau - \lambda)} \left[ f(\tau) - f(\lambda) \right]
	\rightarrow \frac{1}{2} \ {\rm for} \ m_f \to \infty,
\end{equation}
where $\tau = 4 m_f^2 / M_{\phi}^2$, $\lambda = 4 m_f^2 / M_Z^2$, and
\begin{equation}
	f(x) = \left\{ \begin{array}{lc} 
	\left[ \sin^{-1} (1/\sqrt{x}) \right]^2 \ \ & \tau \geq 1, \\
	-\frac{1}{4} \left[ \ln(\eta_+/\eta_-) - i \pi \right]^2 \ \ & \tau < 1,
	\end{array} \right.
\end{equation}
with $\eta_{\pm} \equiv 1 \pm \sqrt{1 - x}$.  Forming the ratio (normalized to 1 at $m_f \to \infty$),
\begin{equation}
	-4 \frac{I_2(\tau,\lambda)}{F_{1/2}^{\phi}(\tau)} = \left\{ \begin{array}{cc}
	1.087 \ \ & {\rm for} \ m_f = 100~{\rm GeV} \\
	1.018 \ \ & {\rm for} \ m_f = 200~{\rm GeV} \end{array} \right.
\end{equation}

This effect can be used to shift the effective $b$ value in the $Z\gamma$ amplitude upward (for a light fermion carrying only weak isospin) or downward (for a light fermion carrying only hypercharge).  We see that the shift in this effective $b$ value is constrained to be less than 9\% in either direction (we take 100~GeV as a lower bound on the mass of such a new fermion due to its non-observation at LEP-2).  Such a shift in the effective value of $b$ is too small to significantly change the prediction for $\Gamma(\phi \to Z \gamma)$, so it does not provide a loophole in the exclusion.

\subsection{Decays of $\phi$ to weak-scale $Z^{\prime}$ boson pairs}

Our prediction of the huge enhancement in the $Z\gamma$ rate for the pseudoscalar rests on the requirement to tune the $\phi \gamma\gamma$ effective coupling to be close to zero in order to match the observed ratio of rates in the $\gamma\gamma$ and $4\ell$ final states.  But what if the observed $4\ell$ final state does not come from the effective vertices in Eq.~(\ref{eqn:phi-L})?  In this case, $b$ is unconstrained and we can have $\Gamma(\phi \to Z\gamma) \sim \Gamma(\phi \to \gamma\gamma)$ as it is for the SM Higgs.

This can be achieved if the observed events in the $4\ell$ final state come from $\phi \to Z^{\prime} Z^{\prime} \to 4 \ell$, for a new neutral gauge boson $Z^{\prime}$ with mass near the $Z$ pole (in order to approximate the kinematics of the SM $H \to ZZ^* \to 4\ell$ decay).  Such a $Z^{\prime}$ under the $Z$ pole is in fact slightly favored by LEP measurements~\cite{Dermisek:2011xu}.  

In this case, a conclusive detection of the $WW^*$ final state at a rate consistent with the SM Higgs would still exclude the pseudoscalar possibility, which predicts $\Gamma(\phi \to WW^*)$ to be more than two orders of magnitude smaller than $\Gamma(\phi \to \gamma\gamma)$ for generic $b$ values not too close to the zero in the $\gamma\gamma$ rate.

%%%%%%%%%%%%%%%%%%%%%%%%%%%%%%%%%%%%%%%
\section{Conclusions}
\label{sec:conclusions}

In this paper we examine the possibility that the newly-discovered 126~GeV boson is a pseudoscalar by parameterizing the couplings to gauge bosons in terms of an effective Lagrangian.  In this framework, the four decay modes to $\gamma\gamma$, $ZZ^*$, $Z\gamma$, and $WW^*$ are controlled by only two effective couplings, i.e., the coefficients of the $\phi B_{\mu\nu} \widetilde B^{\mu\nu}$ and $\phi W_{\mu\nu}^i \widetilde W^{i \mu\nu}$ effective operators.  The ratio of these coefficients can be fixed in terms of the ratio of the highest-significance experimentally-observed rates into $\gamma\gamma$ and $4\ell$, leading to predictions for the rates in $WW^*$ and $Z\gamma$.

An important part of our analysis is the realistic treatment of the $Z\gamma^* \to 4\ell$ contribution to the $4\ell$ final state after experimental cuts.  In fact, we find that $Z\gamma^*$ dominates over $ZZ^*$ by a factor of $\sim 2400$ after cuts.  Taking this into account, the pseudoscalar model predicts a rate $\sigma/\sigma_{\rm SM} \simeq 170$ in the $Z\gamma$ final state relative to the SM Higgs.  While this enhancement is not large enough to show up against the SM $Z\gamma$ total rate, it should be easily visible (or excludable) through a $Z\gamma$ resonance search using only the existing LHC data.  This prediction can be evaded if the observation in the $4\ell$ final state is due to exotic decays of the pseudoscalar to weak-scale $Z^{\prime}$ pairs.

We also find that the decay rate to $WW^*$ for the pseudoscalar is dramatically suppressed compared to that for the SM Higgs, yielding $\sigma/\sigma_{\rm SM} \simeq 1/440$ for this channel in the $e \nu \mu \nu$ final state, which is not contaminated by $Z\gamma^*$ decays.  Conclusive observation of the $WW^*$ final state with a rate near the SM Higgs prediction would thus eliminate the pseudoscalar possibility.  This, however, will require additional integrated luminosity beyond that used already in the Higgs analyses.  The prediction of the suppression of $WW^*$ is robust against new contributions to the $4\ell$ final state from exotic decays of the pseudoscalar to weak-scale $Z^{\prime}$ pairs.

%%%%%%%%%%%%%%%%%%%%%%%%%%%%%%%%%%%%%%%%%
\begin{acknowledgments}
B.C.\ and H.E.L.\ were supported by the Natural Sciences and Engineering Research Council of Canada.  H.E.L.\ thanks S.~Chivukula, S.~Godfrey, J.~Ren and T.~Rizzo for useful conversations. K.K. thanks R. Vega-Morales and S.~Shalgar for helpful discussions.
\end{acknowledgments}

%%%%%%%%%%%%%%%%%%%%%%%%%%%%%%%%%%%%%%%%%%%%%%

\end{document}